\documentclass[a4,11pt]{article}
\pdfoutput=1 
\PassOptionsToPackage{hypertexnames=false}{hyperref}
\usepackage{jheppub}
\usepackage{graphicx}
\usepackage{xcolor, graphics}
\usepackage{amsmath,amssymb,amsfonts}
\usepackage{acronym}
\usepackage{mathrsfs}
 \usepackage{units}
\usepackage{enumerate}   
\usepackage{etoolbox}
\usepackage{float,slashed}
\usepackage{enumitem}
\usepackage{mathtools}
\usepackage{tikz}
\usetikzlibrary{quotes,arrows.meta,calc,patterns,angles}
\usetikzlibrary{%
    decorations.pathreplacing,%
    decorations.pathmorphing%
}
\usetikzlibrary{decorations.markings}
\usetikzlibrary{positioning,arrows,shapes} 
\tikzset{photon/.style={decorate, decoration={snake}, draw=black},
fermion/.style={thick,draw=blue, postaction={decorate},
    decoration={markings,mark=at position .5 with {\arrow[blue]{triangle 45}}}},
gluon/.style={decorate, draw=black,
    decoration={coil,aspect=0}},
scalar/.style={thick,dashed,draw=blue, postaction={decorate}},
higgs/.style={thick,loosely dashed,draw=blue, postaction={decorate}}}    

\newcommand{\gs}{g_\star}
\newcommand{\gss}{g_{\star s}}
\newcommand{\Trh}{T_\text{RH}}

\newcommand{\Mpl}{M_\text{Pl}}

\def\beq{\begin{equation}\begin{aligned}}
\def\eeq{\end{aligned}\end{equation}}

\begin{document}
\title{Constraining the Coexistence of Primordial Black Holes and Particle Dark Matter with Neutrino Observations}
\author[1]{Prolay Chanda,}
\affiliation[1]{Tata Institute of Fundamental Research, Homi Bhabha Road, Mumbai 400005, India}
\author[2]{Sagnik Mukherjee}
\author[2]{and James Unwin}
\affiliation[2]{Department of Physics,  University of Illinois Chicago, Chicago, IL 60607, USA}

\abstract{
Primordial black holes (PBH) with a uniform mass scale could contribute up to 1\% of the gravitationally inferred  dark matter relic abundance and remain consistent with observational limits over a large range of masses. In this case, the vast majority of the dark matter relic abundance is comprised of dark matter particles, such as WIMPs or FIMPs. Particle dark matter gravitationally captured around primordial black holes can form dense minispikes in which the annihilation rate is strongly enhanced. In this work, we investigate the constraints on the coexistence of PBHs and particle dark matter from high-energy neutrino observations. Relative to earlier analyses, we refine the treatment of the dark matter halo profile and its redshift evolution. We consider models of freeze-out and freeze-in dark matter, as well as Boltzmann-suppressed freeze-in. 
We present idealized IceCube event-based sensitivities together with conservative limits obtained by requiring that the predicted extragalactic neutrino intensity not exceed the upper envelope of the measured diffuse flux. We explore the constraints in terms of an idealized model with 100\% branching to neutrinos, we also discuss these results within the context fo a motivated gauged U(1)${}_{L_\mu-L_\tau}$ mediator model, emphasizing that a consistent particle-physics completion generally predicts correlated charged-lepton and neutrino final states.
}
\maketitle

\section{Introduction}

PBHs can form in the early universe if inflation generates a strong enhancement in the primordial curvature power spectrum on small scales \cite{Zeldovich:1967lct,Hawking:1971ei,Carr:1974nx,Carr:1975} (see \cite{Carr:2020xqk} for a review). PBHs formed in this manner can capture the ambient particle dark matter present at PBH formation or dark matter decoupling, whichever occurs later. While the particle dark matter will initially have uniform density, subsequent collapse due to the gravitational pull of the PBH leads to the dark matter halo reassembling into a `spike' profile. Analytic studies of self-similar collapse \cite{Bertschinger:1985pd} in this fashion imply that the halo density profile scales characteristically as $\rho\propto r^{-9/4}$. The central densities of PBH halos are expected to be very large, substantially enhancing annihilation rates. Gamma-ray observations have consequently been used to constrain the coexistence of PBHs with WIMPs \cite{Fermi-LAT:2014ryh,Lacki:2010zf,Adamek:2019gns,Eroshenko:2016yve,Gines:2022qzy,Kadota:2021jhg,Tashiro:2021xnj,Kadota:2022cij,Carr:2020mqm,Scholtz:2019csj,Boucenna:2017ghj,Chanda:2022hls} and, more recently, with FIMPs \cite{Chanda:2025bpl}.

Previous studies have considered the case of velocity-independent cross-section and the resulting neutrino flux \cite{Yang:2013dsa,Yang:2017yjp,Hao:2024hzu}. Moreover, they can provide constraints on dark matter that has no interactions with photons, such as dark matter that only annihilates into neutrinos or cascade annihilations (cf.~\cite{Pospelov:2007mp,Martin:2014sxa,Elor:2015bho}) into hidden sector states that decay to neutrinos.
Dark matter interactions with neutrinos provide a particularly well-motivated framework for probing physics beyond the Standard Model. 

From the particle-physics perspective, neutrinos are the only electrically neutral fermions in the Standard Model and therefore naturally connect to hidden-sector or dark-sector physics through gauge-singlet operators. In addition, neutrino-dark matter interactions can potentially affect cosmological evolution and structure formation. These scenarios have therefore attracted considerable attention across collider, cosmological, and neutrino experiments. In this work, we focus on the possibility that dark matter annihilates into neutrinos, which provides a clean and comparatively model-independent probe of the dark sector while avoiding strong electromagnetic constraints that typically arise in visible annihilation channels.

 Among the possible annihilation channels, the direct process
$\chi\bar\chi \rightarrow \nu\bar{\nu}$ \cite{Arguelles:2019ouk} is especially interesting because it represents one of the most elusive and difficult-to-constrain dark matter signatures. In many conventional dark matter scenarios, annihilations into charged Standard Model particles inevitably generate secondary photons through hadronization, inverse Compton scattering, or final-state radiation, leading to strong gamma-ray constraints. In a phenomenological neutrino-line benchmark, the prompt electromagnetic yield can be small, whereas in  gauge invariant construction, this also produces charged leptons (as we highlight in Section~\ref{sec6}). Furthermore, for non-relativistic dark matter annihilation, the produced neutrinos carry energies approximately equal to the dark matter mass,
$E_\nu \simeq m_\chi,$
resulting in a monochromatic neutrino feature that serves as a distinctive experimental signature~\cite{Yuksel:2007ac}. Modern neutrino observatories such as IceCube, Super-Kamiokande, ANTARES, and KM3NeT probe neutrino energies spanning many orders of magnitude, allowing dark matter annihilation into neutrinos to be constrained over a very broad dark matter mass range.

In the present work, we investigate neutrino signals originating from dark matter annihilation inside primordial black-hole (PBH) induced dark matter spikes and use these signals, in particular IceCube data, to constrain the PBH fraction of dark matter, $f_{\rm PBH}$. Primordial black holes gravitationally enhance the surrounding dark matter density, leading to the formation of compact minihalos or spikes in which the annihilation rate can be significantly amplified relative to the cosmological average~\cite{Boucenna:2017ghj}. Consequently, even a subdominant PBH population may generate an observable neutrino flux through enhanced dark matter annihilation. Since no statistically significant excess consistent with such a signal has been observed in present neutrino data, the resulting neutrino flux limits can be translated into upper bounds on $f_{\rm PBH}$~\cite{Lacki:2010zf}. This approach is particularly appealing because neutrinos can escape dense astrophysical environments essentially unattenuated, allowing them to probe compact annihilation regions that may be inaccessible to electromagnetic observations.

Hao {\em et al.}~\cite{Hao:2024hzu} provided an initial analysis of this signal. Here we revisit the prospect of constraining PBH via high energy neutrino data refining several approximations concerning the PBH-induced minispike profile, source flavor composition, redshift evolution, and employing an approximate IceCube detector model, and quantify how these assumptions affect the inferred limits. 
In particular, our analysis introduces the following refinements and extensions:
\begin{itemize}
\item Rather than adopting the idealized approximation $\rho \propto r^{-9/4}$ for the dark matter profile (as in \cite{Hao:2024hzu}), we model the profile as a set of nested power-laws, the form of which depends on both the mass of the dark matter and the PBH mass.
     \beq
     \rho(r,z)=\left\{
\begin{array}{rcl}
\rho_{\rm core} ,\,\,\,\,\,\,\,\,\,\,\,\,\,\,\,\,\,\,\,\,\,\,\,\,\,\,\,\,\,\,\,\,\,\,\,\,\,\,\,\, &&{r< r_{\rm cut}(z)}\\
\\
\rho_{\rm core}(r/r_{\rm cut}(z))^{-9/4}, && { r_{\rm cut}(z)\leq r<r_{\rm ta}(z_{\rm eq})}\\
\end{array} \right.
\eeq
where $r_{\rm ta}(z)$ is the turnaround scale at redshift $z$
\beq
r_{\rm ta}(z) \approx (2GM_{\rm PBH}t(z)^2)^{1/3}.
     \eeq

Specifically, here we adopt the refined halo profiles derived by Boudaud {\em et al.} \cite{Boudaud:2021irr, Lavalle:2026fhx} and extended to include annihilations in \cite{Chanda:2022hls}.

\item We set bounds on $f_{\rm PBH}$ using IceCube's diffuse extragalactic neutrino data.

\item We extend our study beyond WIMPs to the case of freeze-in dark matter, or {\em FIMPs}.

\end{itemize}

This paper is organized as follows. In Section~\ref{sec2} we review the PBH-induced dark matter density profiles and their modification by annihilations. In Section~\ref{sec3} we derive the extragalactic neutrino intensity. Sections~\ref{sec4} and~\ref{sec5} present event-based projected sensitivities and conservative diffuse-flux constraints, respectively. We outline a motivated UV complete model that could give rise to such signals in Section \ref{sec7}. Then, in Section~\ref{sec8} we adapt these limits to the case of freeze-in dark matter. Section~7 provides concluding remarks.

 \section{Dark Matter minispikes around PBHs}
\label{sec2}

Although $\rho\propto r^{-9/4}$ is a useful characteristic approximation, PBH minispikes generally comprise nested power laws whose structure depends on $m_\chi$, $M_\bullet$, and the dark matter decoupling history \cite{Boudaud:2021irr,Lavalle:2026fhx}. The outer halo scale is the turnaround radius at equality, $r_{\rm eq}\simeq(2GM_\bullet t_{\rm eq}^2)^{1/3}$. 
The scaling of the power-law depends on the mass scale of the PBH. Two characteristic masses, $M_1$ and $M_2$, divide the light, intermediate, and heavy PBH regimes.

For instance, in the intermediate mass PBH regime, the complete density profile $\rho_{\rm grav}$, for collisionless particle dark matter  is given as
\begin{equation}\label{eq:DM-halo-Grav}
    \rho_{\rm grav}(r) \propto \begin{cases}
        r^{-3/4}(r) &~~~~~ r_S<r<r_1 \\
         r^{-3/2}(r) &~~~~~ r_1<r<r_2 \\
        r^{-9/4}(r) &~~~~~ r_2<r<r_{\rm eq} \\
        0 &~~~~~ r_{\rm eq}<r
    \end{cases}~.
\end{equation}
We give the full forms of the halo profiles in the three  PBH mass regimes (light, intermediate, and heavy) in  Appendix \ref{sec:spike-density}.

Dark-matter annihilation effects were incorporated in \cite{Chanda:2022hls}, where it was emphasized that, for WIMP-like candidates, annihilations can substantially reshape the late-time inner part of the profile in Eq.~\eqref{eq:DM-halo-Grav}. (See also recent analytic developments in \cite{Lavalle:2025rnx}.)
 In the presence of dark matter self-annihilations the non-collisional dark matter density reduces to a maximum density profile near the center
\begin{align}
\rho_{\rm max}(z)
&\simeq \frac{m_\chi}{\langle\sigma v\rangle_{\rm tot}[t(z)-t_{\rm in}]},
\end{align}
where $t_{\rm in}$ is the age of the halo. Thus, this form effectively assumes order one annihilation event during the age of the halo.

Therefore, following the depletion of the inner dark matter halo due to annihilations, the modified density profile will be
\beq
\rho_{\rm DM}(r,z) = {\rm min}\left[\rho_{\rm max}(z),\rho_{\rm grav}(r,z)\right].
\eeq
The dark matter annihilation rate inside the halo for a given redshift $z$ is given as 
\beq
\Gamma_{\rm ann}(z)=\xi_f\frac{4\pi}{m_\chi^2}
\int dr\,r^2\langle\sigma v\rangle_{\rm tot}\rho_{\rm DM}^2(r,z),
\eeq
where $\xi_f=1/4$ for symmetric Dirac dark matter and $\xi_f=1/2$ for self-conjugate dark matter. For Dirac dark matter, $\rho_{\rm DM}$ denotes the total mass density of particles plus antiparticles. 

The annihilation rate is a function of redshift since the profile evolves over time due to the depletion of the inner regions via annihilations. The evolution with redshift is of the form
\beq
\Gamma_{\rm ann}(z)\simeq \Gamma_{\rm ann}(0)h^x(z),\qquad h(z)\simeq \frac{H(z)}{H_0},
\eeq
where the exponent $x$ depends on the relative mass of the PBH \cite{Chanda:2022hls}.
For example, $x=2/3$ applies only to a restricted velocity-independent, heavy-PBH regime.

We note that for $p$-wave annihilation, $\langle\sigma v\rangle(r)=b\langle v_{\rm rel}^2(r)\rangle$ and $\langle v^2(r)\rangle\sim GM_\bullet/r$. The annihilation-limited profile is therefore spatially dependent and is not given by a constant $\rho_{\rm max}$, see  \cite{Chanda:2022hls}. Here we restrict our attention to velocity-independent ($s$-wave) annihilations.

\section{Extragalactic Neutrino Flux}
\label{sec3}

In this section, we derive the extragalactic neutrino flux generated by dark-matter annihilation within the PBH halos. We relate the redshift-dependent annihilation rate of an individual halo to the cosmological PBH number density, and then obtain the observed neutrino spectrum after accounting for cosmological redshifting and the flavor composition at Earth.

We first note that the physical number density at redshift $z$ evolves as
\beq
n_{\rm PBH}(z) = n_{\rm PBH,0}(1+z)^3,
\eeq
where the comoving number density of PBHs today is given by
\beq
n_{\rm PBH,0} = \frac{f_{\rm PBH}\Omega_{\rm DM}\rho_{\rm crit,0}}{M_{\bullet}},
\eeq
 $f_{\rm PBH}$ being the fraction of dark matter in PBHs, $\Omega_{\rm DM}$ is the dark matter density parameter, and $\rho_{\rm crit,0} = 3H_0^2/(8\pi G)$ is the present day critical density.

The flavor-resolved neutrino intensity is
\begin{equation}
\frac{d\Phi_{\nu_\beta}}{dE_\nu}
=\frac{1}{4\pi}\sum_\alpha P_{\alpha\beta}
\int_0^{z_{\rm max}}dz
\frac{n_{\rm PBH}(z)\Gamma_{\rm ann}(z)}{H(z)(1+z)^3}
\frac{dN_{\nu_\alpha}}{dE_\nu'},
\label{eq:flavor-flux}
\end{equation}
involving the flavor transition probability matrix
\begin{equation}
P_{\alpha\beta}=\sum_i|U_{\alpha i}|^2|U_{\beta i}|^2,
\end{equation}
and an analogous expression holds for antineutrinos. Here $E_\nu'=E_\nu(1+z)$.

For our numerical calculation, we assume an equal-flavor composition at Earth, i.e.~the neutrino flux flavour ratio is 1:1:1. This approximation is not universal, and given a specific model one should rather calculate the oscillation-averaged source-to-Earth flavor transition. Notably, most choices of the source neutrino ratios would change the muon-neutrino normalization at Earth by $\mathcal{O}(1)$ factor and would not impact the conclusions \cite{Cirelli:2005gh,Asai:2020qlp}.

Taking this equal-flavor benchmark Eq.~\eqref{eq:flavor-flux} becomes
\begin{equation}
\frac{d\Phi_{\nu_\mu+\bar\nu_\mu}}{dE_\nu}
=\frac{1}{4\pi}
\frac{f_{\rm PBH}\Omega_{\rm DM}\rho_{{\rm crit},0}}{3M_\bullet}
\int_0^{z_{\rm max}}\frac{dz}{H(z)}
\Gamma_{\rm ann}(z)
\frac{dN_{\nu+\bar\nu}}{dE_\nu'}.
\label{eq:dPhi-Nu-dE}
\end{equation}
The integration runs over redshift with  $z_{\rm max}=z_{\rm eq}$. The redshift integral extends to $z=0$; galaxy formation does not impose a nonzero lower limit. 

For the direct process $\chi\bar\chi\to\nu_\alpha\bar\nu_\alpha$, the differential neutrino spectrum produced per annihilation is denoted as
\begin{align}
\frac{dN_{\nu_\alpha}}{dE_\nu'}=\delta(E_\nu'-m_\chi),\qquad
\frac{dN_{\bar\nu_\alpha}}{dE_\nu'}=\delta(E_\nu'-m_\chi).
\end{align}
We keep neutrino and antineutrino spectra separate because their charged-current interaction cross sections and detector effective areas are different. Their sum contains two particles per annihilation.

Using $z_\star=m_\chi/E_\nu-1$, the spectra can equivalently be written as
\beq
\frac{dN_{\nu_\alpha}}{dE_\nu'}&=\frac{1}{E_\nu}\delta(z-z_\star),\qquad
\frac{dN_{\bar\nu_\alpha}}{dE_\nu'}&=\frac{1}{E_\nu}\delta(z-z_\star),\qquad
z_\star&=\frac{m_\chi}{E_\nu}-1.
\eeq
For the equal-flavor benchmark, the summed muon-flavor intensity is therefore
\begin{equation}
\frac{d\Phi_{\nu_\mu+\bar\nu_\mu}}{dE_\nu}
=\frac{1}{4\pi}\frac{2}{E_\nu}
\frac{f_{\rm PBH}\Omega_{\rm DM}\rho_{{\rm crit},0}}
{3M_\bullet}
\frac{\Gamma_{\rm ann}(z_\star)}{H(z_\star)}
\Theta(z_\star)\Theta(z_{\rm max}-z_\star).
\label{eq:dPhi-Nu-dE-simple}
\end{equation}
Note that the two Heaviside functions set the right-hand side to zero unless
$0\leq z_\star \leq z_{\rm max}$, or equivalently unless the observed neutrino energy lies in the interval
$\frac{m_\chi}{1+z_{\rm max}}\leq E_\nu \leq m_\chi$.
Thus, the upper endpoint corresponds to annihilations occurring locally at \(z_\star=0\), while the lower endpoint is set by the maximum redshift included in the flux integral. 

\section{Projected limits on $f_{\rm PBH}$ at IceCube}
\label{sec4}

In this section we estimate contained and upward-going muon signals above neutrino backgrounds.
We use the predicted extragalactic neutrino flux generated by dark-matter annihilation in primordial black-hole (PBH) mini-spikes to examine constraints and projected limits on the PBH  fraction $f_{\rm PBH}$. Two complementary methods are employed:
\begin{enumerate}
\item Flux-based comparison using the measured diffuse neutrino spectrum. 
\item Event-based comparison using the expected number of detected neutrino events as compared to the background.
\end{enumerate}
The first method is conservative and spectrum-oriented, while the second is detector-oriented and directly connected to counting statistics.

\begin{figure}[t]
\centerline{ \includegraphics[scale = 0.5]{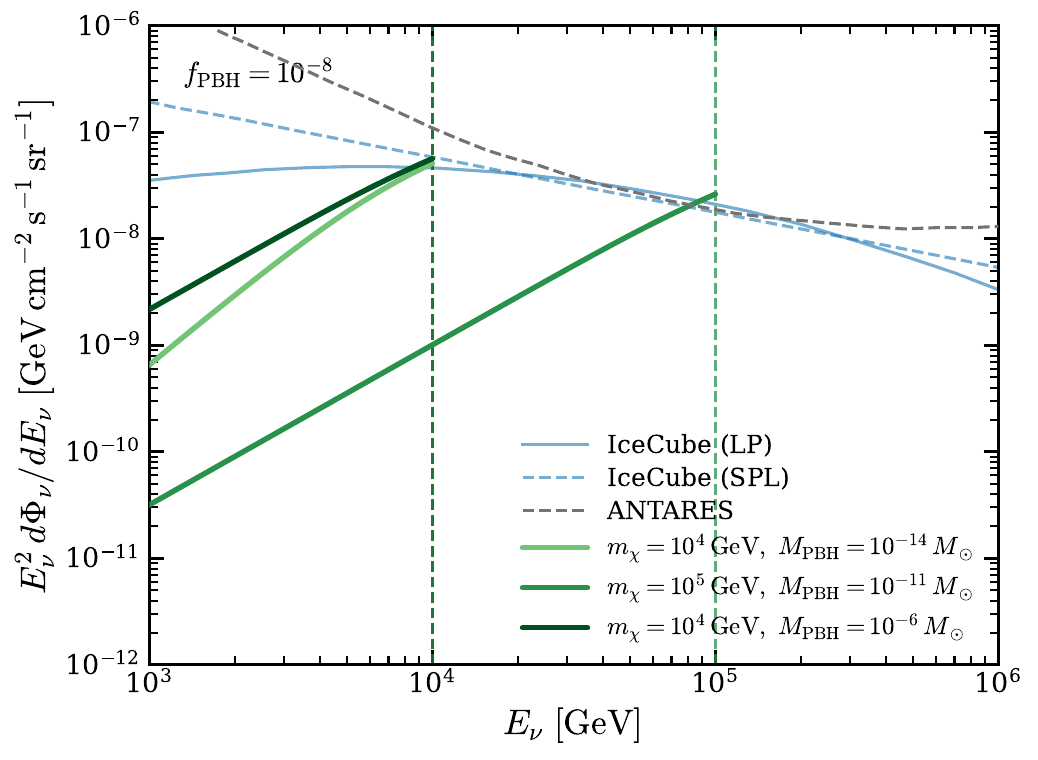}}
   \caption{Extragalactic neutrino flux from dark matter annihilation in the halo around PBHs for representative dark matter masses, compared with the diffuse neutrino flux measurements from IceCube~\cite{Abbasi:2021qfz,IceCube:2025ewu}, and ANTARES \cite{ANTARES:2024ihw}. The theoretical flux scales linearly with the primordial black hole fraction, $f_{\rm PBH}$, and the observational upper limits on $f_{\rm PBH}$ are obtained by requiring that the predicted flux does not exceed the measured diffuse neutrino flux over the energy range considered. The log-parabolic (LP) and single-power-law (SPL) parameterizations are taken from Abbasi~\textit{et al.}~\cite{IceCube:2025ewu}. }
\label{Fig0}\end{figure}

Figure~\ref{Fig0} shows the predicted extragalactic $\nu_\mu+\bar\nu_\mu$ flux for several representative dark matter masses, together with the diffuse neutrino flux measured by IceCube and ANTARES. The comparison illustrates the conservative envelope method in which the predicted signal is required not to exceed the observed diffuse flux. Henceforth, we use the IceCube log-parabolic (LP) fit as the reference model for the diffuse astrophysical neutrino flux when deriving observational constraints.

\subsection{Muon flux from extragalactic neutrinos}

As the extragalactic neutrino flux reaches terrestrial detectors, neutrinos may produce observable muons through charged-current interactions with matter \cite{Erkoca:2009by,Bergstrom:1997tp}. Two classes of events are commonly distinguished. In contained events the neutrino interaction occurs inside the instrumented detector volume. In upward-going events the neutrino traverses the Earth before interacting in the surrounding rock or ice, producing a muon that subsequently enters the detector.

For upward events, neutrinos interact outside the detector and the produced muons propagate through the medium before reaching the detector. The corresponding flux is
\begin{align}
\frac{d\Phi_\mu^{\rm up}}{dE_\mu}
&=\frac{N_A\rho_{\rm md}}{2}\int_{E_\mu}^{m_\chi}dE_\nu
\Bigg[
\frac{d\Phi_{\nu_\mu}}{dE_\nu}\mathcal T_\nu
\left(\frac{d\sigma_{\nu p}}{dE_\mu}+\frac{d\sigma_{\nu n}}{dE_\mu}\right)
\nonumber\\
&\hspace{4cm}+
\frac{d\Phi_{\bar\nu_\mu}}{dE_\nu}\mathcal T_{\bar\nu}
\left(\frac{d\sigma_{\bar\nu p}}{dE_\mu}+\frac{d\sigma_{\bar\nu n}}{dE_\mu}\right)
\Bigg]R_\mu(E_\mu),
\label{eq:upward-muon}
\end{align}
involving transmission probability
\begin{equation}
\mathcal T_\nu(E_\nu,\theta_z)=
\exp\!\left[-N_AX(\theta_z)\left(\sigma_{\nu N}^{\rm CC}(E_\nu)+\sigma_{\nu N}^{\rm NC}(E_\nu)\right)\right].
\label{eq:earth-transmission}
\end{equation}
Here we approximate $\mathcal T_\nu=\mathcal T_{\bar\nu}=1$.
The muon range $R(E_\mu)$ accounts for energy loss in the medium and is given by
\beq
R(E_\mu)=\frac{1}{\rho_{\rm md} \beta}
\ln\left(\frac{\alpha + \beta E_\mu}{\alpha + \beta E_\mu^{\rm th}}\right).
\eeq
Here $\alpha\sim 10^{-3}~{\rm GeV\,cm^{2}\,g^{-1}}$ and $\beta\sim 10^{-6}~{\rm cm^{2}\,g^{-1}}$ parameterize ionization and radiative energy losses, respectively, and $E_\mu^{\rm th}$ is the detector threshold energy.

\begin{figure}[t]
\includegraphics[width=0.45\textwidth]{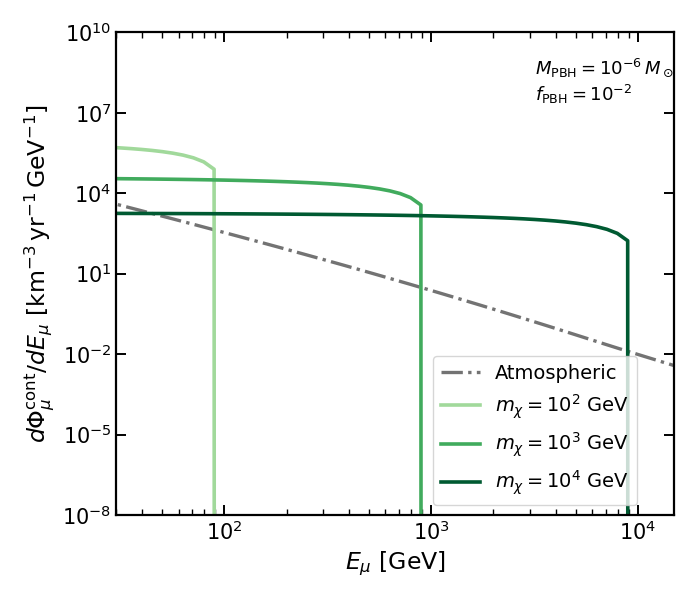}
\includegraphics[width=0.45\textwidth]{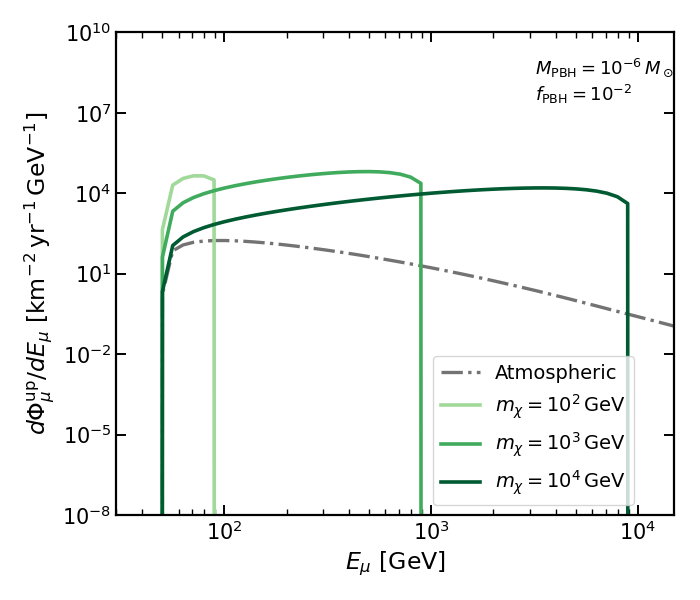}
\vspace{-6mm}
\caption{Differential muon flux for contained events (left) and upward-going events (right) arising from the extragalactic $\nu_\mu+\bar\nu_\mu$ intensity produced by dark matter annihilation in PBH minispikes. Results are shown for three dark matter masses. At higher masses the atmospheric background \cite{Honda:2015fha} falls at large energies. The endpoint is redshift broadened and terminates near $E_\mu\simeq m_\chi$.}
\label{Fig1}
\end{figure}

The muon flux is converted into the number of detected events by integrating over the detector response
\beq
N_{\nu_\mu,{\rm PBH}}=\int_{E_\mu^{\rm th}}^{E_\mu^{\rm max}}dE_\mu\,\frac{d\phi_\mu}{dE_\mu}\,F_{\rm eff}(E_\mu),
\eeq
where $F_{\rm eff}(E_\mu)$ denotes the effective volume for contained events or the effective area for upward events. We take the threshold value to be $E_\mu^{\rm th}\simeq50$ GeV. We compare the dark matter flux to the background atmospheric flux, where we use the background estimates of Honda \textit{et al.} \cite{Honda:2015fha}.

Figure~\ref{Fig1} shows the predicted differential muon flux for contained (left) and upward-going (right) events arising from the extragalactic $\nu_\mu+\bar\nu_\mu$ flux produced by dark matter annihilation in PBH-induced minispikes, for three representative dark matter masses. As the dark matter mass increases, the signal extends to higher energies where the atmospheric neutrino background falls rapidly \cite{Honda:2015fha}, improving the signal-to-background contrast. In each case, the sharp monochromatic neutrino injection at production is broadened by cosmological redshift, producing a characteristic spectral endpoint that terminates near $E_\mu \simeq m_\chi$.

%%%%%%%%%%%%%%%%%%%%%%%%%%%%%%%%%%%%%

\subsection{Projected sensitivity  and $f_{\rm PBH}$ limit from atmospheric neutrino background}

Unlike the observational limits obtained by comparing the predicted neutrino flux with the measured diffuse IceCube flux, our projected sensitivity is determined from the expected atmospheric neutrino background. We first compute the muon event rates induced by the PBH-generated neutrino flux and then determine the value of the PBH abundance for which the signal becomes statistically distinguishable from the atmospheric background.

The contained-muon flux is obtained from the incident neutrino flux through charged-current interactions within the detector volume following the standard treatment of \cite{Bergstrom:1997tp,Erkoca:2009by}. In the idealized detector model adopted here, the resulting contained-muon flux is integrated over the effective detector volume to estimate the event rate. For contained events, the total number of detected muons is given by
\begin{align}
N_{\rm cont}=T_{\rm exp}\,V_{\rm eff}\int dE_\mu\,\frac{d\Phi_\mu^{\rm cont}}{dE_\mu},
\label{eq:Ncontained}
\end{align}
where $T_{\rm exp}$ is the detector exposure time and $V_{\rm eff}$ denotes the effective detector volume. Here take $T_{\rm exp}=10$ years and $V_{\rm eff}=0.04 ~{\rm km}^3$.

For the idealized upward-muon assumed here, the expected number of events reduces to
\begin{equation}
\frac{dN^{\rm up}}{dE_\mu}
=T_{\rm exp}\int_{0}^{1}d(\cos\theta)\,
\frac{d\Phi_\mu^{\rm up}}{dE_\mu d\Omega}
A_{\rm eff}(E_\mu,\theta),
\label{eq:dNdE_master}
\end{equation}
where $A_{\rm eff}(E_\mu,\theta)$ is the approximate IceCube effective area and $d\Phi_\mu^{\rm up}/dE_\mu d\Omega$ is the differential upward-going muon flux at the detector. Throughout this work we employ the following parameterization of the detector response
\beq
A_{\rm eff}(E_\mu,\theta)\simeq2\pi\,A_0(E_\mu)\left(0.92-0.45\cos\theta\right),
\label{eq:Aeff_fit}
\eeq
where $A_0(E_\mu)$ has dimensions of area and is parameterized as
\beq
A_0(E_\mu)=\begin{cases}
0,&E_\mu\le10^{1.6}\ {\rm GeV},\\
0.748\left[\log_{10}\!\left(\frac{E_\mu}{\rm GeV}\right)-1.6\right]{\rm km}^2,&10^{1.6}\le E_\mu\le10^{2.8}\ {\rm GeV},\\
\left[0.9+0.54\left(\log_{10}\!\left(\frac{E_\mu}{\rm GeV}\right)-2.8\right)\right]{\rm km}^2,&E_\mu\ge10^{2.8}\ {\rm GeV}.
\end{cases}
\label{eq:A0_piecewise}
\eeq
This provides the approximate detector response used throughout the numerical analysis.

Since the neutrino flux from PBH-induced dark matter annihilation is computed assuming $f_{\rm PBH}=1$, the signal event number scales linearly with the PBH abundance
\beq
N_{\rm PBH}(f_{\rm PBH})=f_{\rm PBH}\,N_{\rm PBH}^{(f=1)},
\eeq
where $N_{\rm PBH}^{(f=1)}$ is the event number calculated for the reference choice $f_{\rm PBH}=1$.

To estimate the projected sensitivity, we define the statistical significance as
\beq
\zeta=\frac{N_{\rm PBH}}{\sqrt{N_{\rm PBH}+N_{\rm ATM}}}.
\label{eq:significance}
\eeq
Throughout this work we adopt a projected $2\sigma$ sensitivity, corresponding to
\beq
\zeta_{\rm crit}=2,
\eeq
which approximately corresponds to the signal producing a two-standard-deviation excess above the expected atmospheric background under Gaussian counting statistics.

Imposing
$\zeta=\zeta_{\rm crit},$
yields the projected upper limit on the PBH dark matter fraction,
\beq
f_{\rm PBH}^{\rm max}=\frac{\zeta_{\rm crit}^{\,2}+\zeta_{\rm crit}\sqrt{\zeta_{\rm crit}^{\,2}+4N_{\rm ATM}}}{2N_{\rm PBH}^{(f=1)}}.\label{eq:fPBHlimit}
\eeq
The atmospheric neutrino flux provides the irreducible background. In general, heavier dark matter produces more energetic neutrinos and hence longer-ranged muons, leading to improved sensitivity, particularly for upward-going events where the effective detection area increases rapidly with muon energy.

\begin{figure}[t]
\includegraphics[width=0.5\textwidth]{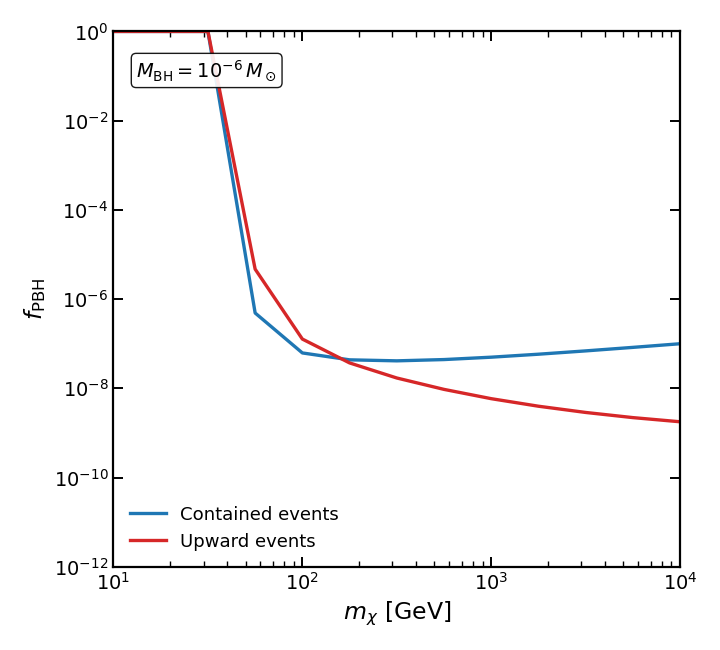}
\includegraphics[width=0.5\textwidth]{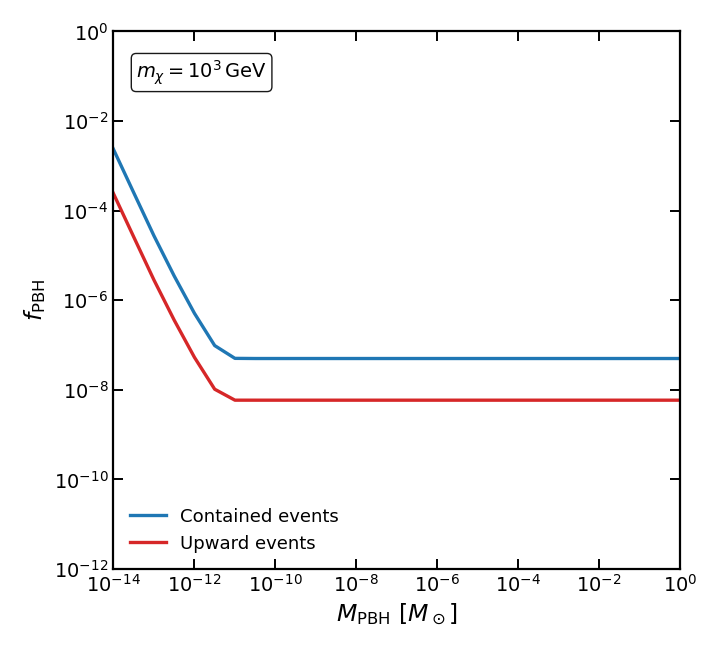}
\vspace{-3mm}   \caption{Idealized IceCube projected sensitivities to $f_{\rm PBH}$ in the mixed WIMP-PBH scenario. Left: dependence on $m_\chi$ for $M_\bullet=10^{-6}M_\odot$. Right: dependence on $M_\bullet$ for $m_\chi=10^3~{\rm GeV}$.}
\label{fig:fo-event}
\end{figure}

%%%%%%%%%%%%%%%%%%%%%%%%%%%%%%%%%%%%%

In Figure~\ref{fig:fo-event}, we present the resulting idealized IceCube projected sensitivities to the PBH fraction obtained from the contained and upward-going event analyses. For fixed $M_{\rm PBH}=10^{-6}\,M_\odot$, the sensitivity improves rapidly with increasing dark matter mass before saturating once the atmospheric background becomes subdominant. Likewise, for fixed $m_\chi=10^3\,\mathrm{GeV}$, the projected sensitivity becomes largely independent of $M_{\rm PBH}$ over a broad mass range, reflecting the weak dependence of the annihilation luminosity on the black-hole mass once the minispike structure has reached its asymptotic scaling regime. In Appendix  \ref{ApD} we discuss how the three scaling regimes in the $f_{\rm PBH}$-$m_\chi$ limits arise.

\section{Limits on \texorpdfstring{$f_{\rm PBH}$}{fPBH} from the observed extragalactic neutrino flux}
\label{sec5}

To constrain the PBH fraction, we compare the predicted diffuse $\nu_\mu+\bar\nu_\mu$ intensity with diffuse-flux measurements from IceCube~\cite{Abbasi:2021qfz,IceCube:2025ewu} and ANTARES~\cite{ANTARES:2024ihw}. The IceCube single-power-law and log-parabolic spectra are taken from \cite{IceCube:2025ewu}. For fixed $m_\chi$ and $M_\bullet$, we compute the coefficient of the term linear in $f_{\rm PBH}$. Since the comoving PBH number density scales linearly with $f_{\rm PBH}$
\beq
n_{\rm PBH}\propto f_{\rm PBH},
\eeq
the predicted neutrino flux also scales linearly
\beq
E_\nu^2\frac{d\Phi_{\nu_\mu+\bar\nu_\mu}^{\rm th}}{dE_\nu}
=f_{\rm PBH}\left[E_\nu^2\frac{d\Phi_{\nu_\mu+\bar\nu_\mu}^{\rm th}}{dE_\nu}\right].
\eeq

The prediction is converted to the same flavor, solid-angle, and neutrino-plus-antineutrino convention as the observational result. In particular, a track-based $\nu_\mu+\bar\nu_\mu$ measurement is not compared with an all-flavor flux. We use the energy interval in which both theory and data are available
\beq
E_{\rm min}&=\max\!\left(E_{\rm obs}^{\rm min},\frac{m_\chi}{1+z_{\rm max}}\right),\\
E_{\rm max}&=\min\!\left(E_{\rm obs}^{\rm max},m_\chi\right).
\eeq
At every energy in this interval, the conservative envelope condition is
\beq
f_{\rm PBH}\left[E_\nu^2\frac{d\Phi_{\nu_\mu+\bar\nu_\mu}^{\rm th}}{dE_\nu}\right]
\leq\left[E_\nu^2\frac{d\Phi_{\nu_\mu+\bar\nu_\mu}^{\rm obs}}{dE_\nu}\right]_{\rm upper},
\eeq
where the right-hand side is the published upper uncertainty envelope in the relevant energy bin. This yields
\beq
f_{\rm PBH}^{\rm lim}(E_\nu)=
\frac{\left[E_\nu^2 d\Phi_{\nu_\mu+\bar\nu_\mu}^{\rm obs}/dE_\nu\right]_{\rm upper}}
{\left[E_\nu^2 d\Phi_{\nu_\mu+\bar\nu_\mu}^{\rm th}/dE_\nu\right]}~.
\eeq
The final constraint is taken to be the most restrictive value over the
accessible energy range,
\beq
f_{\rm PBH}^{\rm lim}=\min\!\left[1,\,\min_{E_\nu\in[E_{\rm min},E_{\rm max}]}
f_{\rm PBH}^{\rm lim}(E_\nu)\right],
\eeq
where the outer minimum simply enforces the physical requirement
$f_{\rm PBH}\le1$.

The pointwise condition discards correlations between energy bins and does not exploit the redshift-broadened endpoint. It is therefore a conservative envelope bound, not an optimized spectral limit. A spectral analysis instead uses
\begin{equation}
\Phi_{\rm model}(E_\nu)=\Phi_{\rm astro}^{\rm SPL/LP}(E_\nu)
+f_{\rm PBH}\Phi_{\rm PBH}(E_\nu),
\end{equation}
and profiles the astrophysical normalization and spectral parameters together with $f_{\rm PBH}$. Such a fit is required for an optimized constraint; it is not encoded in the supplied pointwise-bound curve.

\begin{figure}
\centerline{ \includegraphics[scale = 0.47]{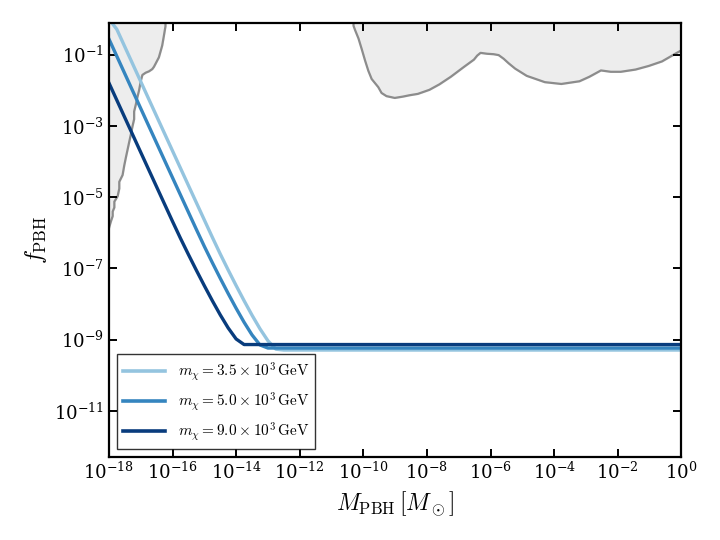}}
   \caption{Conservative diffuse-flux envelope bound on $f_{\rm PBH}$ for the phenomenological $\chi\bar\chi\to\nu\bar\nu$ channel. The prediction is matched to the observed $\nu_\mu+\bar\nu_\mu$ and solid-angle convention. The grey shaded region shows the range of $M_{\rm PBH}$ and $f_{\rm PBH}$ already ruled out by current observational constraints, including Hawking evaporation, microlensing, gravitational waves, and CMB distortions (see, e.g.~\cite{Carr:2020xqk}).}
      \label{fig:fo-flux}
\end{figure}

In Figure~\ref{fig:fo-flux}, we present the conservative diffuse-flux envelope bound obtained by requiring that the predicted extragalactic $\nu_\mu+\bar{\nu}_\mu$ intensity from PBH minispikes does not exceed the observed diffuse neutrino flux. Unlike the event-based sensitivities of Fig.~\ref{fig:fo-event}, these limits rely directly on the measured spectrum and therefore provide a robust, detector-independent constraint. The projected bound is strongest for low PBH masses, where the larger PBH number density enhances the cumulative annihilation signal, before approaching a broad plateau once the minispike luminosity compensates for the declining PBH abundance.

\section{Example Model for Dark Matter Annihilation into Neutrinos}
\label{sec6}

In this section, we present an illustrative anomaly-free vector-mediator completion together with the corresponding low-energy effective interaction in which dark matter annihilation into neutrinos can constitute an significant branching fraction. It should be emphasized that simplified models in which dark matter annihilates exclusively through the process $\chi\bar\chi\rightarrow\nu\bar\nu$ are generally not compatible with the Standard Model gauge structure. Since the left-handed neutrinos reside in SU(2)$_L$ lepton doublets, an interaction that singles out neutrinos alone violates electroweak gauge invariance unless accompanied by additional structure. Consequently, ultraviolet-complete realizations inevitably generate correlated interactions involving the associated charged leptons.

As a minimal proof of principle, we therefore consider a leptophilic $Z'$ arising from an anomaly-free gauged U(1)${}_{L_\mu-L_\tau}$ symmetry. Although this model predicts additional charged-lepton annihilation channels and the associated multi-messenger phenomenology, it provides a simple example in which a sizeable fraction of dark matter annihilations proceed into neutrinos. Our analysis in the main text remains intentionally phenomenological, adopting a neutrino benchmark to isolate the neutrino signal, while the explicit model illustrates how such a benchmark may be embedded within a consistent gauge theory. Since the resulting neutrino constraints are expected to differ only at the $\mathcal{O}(1)$ level after accounting for the altered branching fractions and flavor composition, we defer a comprehensive model-dependent analysis of the correlated neutrino and charged-lepton signatures to future work.

\subsection{Anomaly-free leptophilic vector mediator}
\label{app:LmuLtau_model}

For definiteness, and to avoid naturalness issues, we take the dark matter to be a Dirac fermion $\chi$. The dark matter is coupled to a  leptophilic $Z'$ with relevant interaction terms
\begin{equation}
    \mathcal L
    \supset
    Z'_\mu \bar\chi \gamma^\mu
    \left(g' q_\chi ^V - g' q_\chi ^A \gamma^5 \right)\chi
    +
    g_L Z'_\mu \bar L \gamma^\mu L .
    \label{eq:naive_Zprime_L}
\end{equation}
Since $L$ is the Standard Model lepton doublet, it involves both an active lefthanded (LH) neutrino component $\nu_{iL}$ and a Standard Model lepton $l_{iL}$ implying $Z'$ couplings to both states
\begin{equation}
  g_L Z'_\mu     \bar L_i \gamma^\mu L_i
    =
  g_L Z'_\mu     \bar\nu_{iL}\gamma^\mu \nu_{iL}
    +
  g_L Z'_\mu     \bar l_{iL}\gamma^\mu l_{iL}.
\end{equation}
Thus a neutrino-only coupling is not electroweak invariant: a coupling to $L_i$ necessarily includes the charged lepton in the same doublet.

Moreover, a new gauge boson coupled to $\bar L_i\gamma^\mu L_i$ is not automatically associated with an anomaly-free gauge symmetry.  Notably,  U(1)${}_{L_\mu-L_\tau}$ is automatically anomaly-free within the unadorned Standard Model. Thus, to avoid gauge anomalies (or the necessity of additional {\em ad hoc} fields purely for the purpose of anomaly cancelation), we consider gauged U(1)${}_{L_\mu-L_\tau}$ as a concrete motivated example. Moreover, gauged $L_\mu-L_\tau$ has been studied in the literature as an extension of the Standard Model with independent motivation within the context of dark matter and neutrino phenomenology~\cite{Baek:2008nz,Asai:2020qlp,He:1991qd,He:1990pn,Choubey:2004hn}.

For the dark matter to couple to the $Z'$ it must carry a non-zero U(1)${}_{L_\mu-L_\tau}$  charge $q_\chi$. Since the dark matter is a fermion, to avoid introducing an anomaly we introduce a vector-like pair of Weyl fermions, or equivalently we require that the Dirac fermion $\chi$ is such that its LH and RH components have equal charges.  It follows that the relevant interactions are
\beq
    \mathcal L
    \supset&
    -\frac14 Z'_{\mu\nu}Z'^{\mu\nu}
    +\frac12 m_{Z'}^2 Z'_\mu Z'^\mu
    +
    \bar\chi (i\slashed D-m_\chi)\chi
    +
    g' Z'_\mu J^\mu_{L_\mu-L_\tau}
    +
    g' q_\chi  Z'_\mu \bar\chi\gamma^\mu\chi .
    \label{1}
\eeq
The $Z'$ mass may arise either from a dark Higgs field charged under U(1)${}_{L_\mu-L_\tau}$ or from a Stueckelberg mechanism \cite{Stueckelberg:1938hvi}.  At energies below the symmetry-breaking scale, Eq.~\eqref{1} is a valid effective theory.

\subsection{Annihilation into neutrinos and charged leptons}

In the $L_\mu-L_\tau$ model,
\begin{equation}
    g_{\nu_\mu}=+g',
    \qquad
    g_{\nu_\tau}=-g',
    \qquad
    g_{\nu_e}=0 .
\end{equation}
The annihilation cross section for Dirac dark matter via vector coupling to $Z'$ into a single massless active neutrino flavor is
\begin{equation}
    \sigma v(\chi\bar\chi\to \nu_i\bar\nu_i)
 \simeq
\frac{q_\chi^2g'^4}{2\pi}
\frac{m_\chi^2}
{\left(4m_\chi^2-m_{Z'}^2\right)^2
+m_{Z'}^2\Gamma_{Z'}^2},
    \qquad i=\mu,\tau .
\end{equation}
Therefore, summing over $\nu_\mu$ and $\nu_\tau$
\begin{equation}
    \sigma v(\chi\bar\chi\to \nu\bar\nu)
    \simeq
    \frac{q_\chi^2 g'^4}{\pi}
    \frac{m_\chi^2}
    {  \left(4m_\chi^2-m_{Z'}^2\right)^2+m_{Z'}^2\Gamma_{Z'}^2} .
\end{equation}
For \(m_{Z'}\gg m_\chi\), this reduces to
\begin{equation}
    \sigma v(\chi\bar\chi\to \nu\bar\nu)
    \simeq
     \frac{q_\chi^2 g'^4}{\pi}
    \frac{m_\chi^2}{m_{Z'}^4}.
\end{equation}

Since the $Z'$ couples to the muon and tau doublets, and our regime of interest is $m_\chi\gg m_\mu, m_\tau$, the dark matter can also annihilate via
$\chi\bar\chi\to \mu^+\mu^-$, and $\chi\bar\chi\to \tau^+\tau^-$.
The charged-lepton annihilation rate for a final-state charged lepton
\(l=\mu,\tau\) is approximately
\begin{equation}
    \sigma v(\chi\bar\chi\to l^+l^-)
    \simeq
     \frac{q_\chi^2 g'^4}{\pi}
    \frac{m_\chi^2}
    {\left(4m_\chi^2-m_{Z'}^2\right)^2+m_{Z'}^2\Gamma_{Z'}^2}
    \left(1+\frac{m_l^2}{2m_\chi^2}\right)
    \sqrt{1-\frac{m_l^2}{m_\chi^2}} .
\end{equation}
For \(m_l\ll m_\chi\ll m_{Z'}\), this reduces to
\begin{equation}
    \sigma v(\chi\bar\chi\to l^+l^-)
    \simeq
     \frac{ q_\chi^2 g'^4}{\pi}
    \frac{m_\chi^2}{m_{Z'}^4}.
\end{equation}
In this limit the branching fractions are
\begin{align}
{\rm Br}(\chi\bar\chi\to\nu\bar\nu)&\simeq\frac{1}{3},\\
{\rm Br}(\chi\bar\chi\to\mu^+\mu^-)&\simeq\frac{1}{3},\\
{\rm Br}(\chi\bar\chi\to\tau^+\tau^-)&\simeq\frac{1}{3},
\end{align}
for $m_\chi\gg m_\tau$ and $m_{Z'}\gg m_\chi$. Thus approximately two thirds of the annihilations enter charged-lepton final states, and correlated neutrino and gamma-ray signals are unavoidable.
The flavor composition of this explicit model is not used in the numerics of the main body, which employ the equal-flavor approximation. Implementing the model-specific oscillated composition would change the muon-neutrino normalization at an order-unity level but would not alter the qualitative conclusions \cite{Cirelli:2005gh,Asai:2020qlp}.

In the above we have assumed that $ m_{Z'}^2\Gamma_{Z'}^2 \ll m_{Z'}^4$, and we should verify this assumption. In the heavy-mediator regime \(m_{Z'}\gg m_\chi,m_\tau\), the mediator can decay into both Standard Model leptons and dark matter.  The Standard Model contribution is
\begin{equation}
    \Gamma_{Z'}^{\rm SM}= 2\,\frac{g'^2m_{Z'}}{24\pi}+2\,\frac{g'^2m_{Z'}}{12\pi}= \frac{g'^2m_{Z'}}{4\pi},
\end{equation}
where the first term comes from \(\nu_\mu,\nu_\tau\) and the second from
\(\mu,\tau\).  The dark matter contribution is
\begin{equation}
    \Gamma_{Z'}^{\rm DM}
    =
    \frac{q_\chi^2 g'^2 m_{Z'}}{12\pi}
    \left(1+\frac{2m_\chi^2}{m_{Z'}^2}\right)
    \sqrt{1-\frac{4m_\chi^2}{m_{Z'}^2}} .
\end{equation}
Thus, for \(m_{Z'}\gg m_\chi,m_\tau\),
\begin{equation}
    \Gamma_{Z'}=  \Gamma_{Z'}^{\rm SM}+  \Gamma_{Z'}^{\rm DM}
    \simeq
    \frac{g'^2m_{Z'}}{12\pi}
    \left(3+q_\chi^2\right).
\end{equation}
The approximation \(m_{Z'}^2\Gamma_{Z'}^2\ll m_{Z'}^4\) is then equivalent to
\(\Gamma_{Z'}/m_{Z'}\ll 1\), and thus holds for typical models.

\subsection{Annihilation into photons}

For the pure-vector interaction considered here, the one-loop triangle amplitude coupling one $Z'$ current to two electromagnetic currents vanishes by charge conjugation. Consequently, the process $\chi\bar\chi\to Z'^{*}\to\gamma\gamma$ is absent at this order, including for an off-shell \(Z'\). The Landau-Yang theorem independently forbids an on-shell spin-one mediator from decaying into two on-shell photons   \cite{Landau:1948kw,Yang:1950rg} (cf.~also \cite{Keung:2008ve}).  
Accordingly, the leading $\gamma$-ray channel relevant for indirect detection in the leptophilic model for $m_\chi\gg m_\tau$ is the annihilation channel $\chi\bar\chi\to \tau^+\tau^-.$ 
We note that although \(\chi\bar\chi\to\mu^+\mu^-\) and \(\chi\bar\chi\to\tau^+\tau^-\) have comparable tree-level rates, the prompt gamma-ray signal is dominated by the \(\tau^+\tau^-\) channel because tau decays frequently produce neutral pions, \(\pi^0\to\gamma\gamma\), whereas muon final states yield photons only through radiative corrections and secondary processes \cite{Cirelli:2010xx}.

To derive the bounds on dark matter annihilations, one uses the photon spectrum due to annihilations to $\tau^+\tau^-$, whose decays have been characterized by Monte Carlo, see e.g.~\cite{Cirelli:2010xx}.
Thus to compute the $\gamma$-ray limits for PBH minispikes one uses the differential flux
\begin{equation}
    \left.
    \frac{d\Phi_\gamma}{dE\,d\Omega}
    \right|_{\rm ExGal}
    =
    \int_0^\infty dz\,
    \frac{\widehat{\Gamma}_\bullet(z)\, n_{\rm PBH}}
    {8\pi H(z)}
    e^{-\tau(z,E')}
   \left[ \frac{dN_\gamma}{dE'}\right]_\tau.
\end{equation}
with the rate 
\begin{equation}
    \Gamma_\bullet
    =
    4\pi
    \int dr\, r^2
    \left(
        \frac{\rho(r)}{2m_\chi}
    \right)^2
\left[    \langle \sigma v\rangle \right]_\tau.
\end{equation}
Taking the cross section and $\gamma$-ray spectrum $dN_\gamma/dE'$ (from \cite{Cirelli:2010xx}) appropriate to $\tau$ pair production and subsequent decay.

 As expected, the explicit U(1)$_{L_\mu-L_\tau}$ realisation leads to both neutrino and charged-lepton final states, this reducing the neutrino branching fraction relative to the phenomenological benchmark adopted in the main text. At the same time, however, the charged-lepton channels generate correlated electromagnetic signatures, opening the possibility of complementary gamma-ray observables and thus multi-messenger probes. A dedicated study combining neutrino and gamma-ray observations within this explicit framework is left for future work.

\section{Limits on Freeze In Dark Matter Around PBH}
\label{sec7}

In contrast to freeze-out, in freeze-in dark matter initially has a negligible abundance and is gradually produced from thermal bath interactions \cite{Hall:2009bx,Elahi:2014fsa}. We use the dimension-six interaction associated with the heavy-mediator limit of Section~\ref{sec6}, specifically
\begin{equation}
\mathcal L_{\rm eff}=\frac{1}{\Lambda^2}(\bar\chi\gamma_\mu\chi)
\left[\bar L_\mu\gamma^\mu L_\mu+\bar\mu_R\gamma^\mu\mu_R
-\bar L_\tau\gamma^\mu L_\tau-\bar\tau_R\gamma^\mu\tau_R\right].
\label{eq:freezein-operator}
\end{equation}
For an operator of mass dimension $6$, the freeze-in yield has the parametric form
\beq \label{eq:Yafter-app}
    Y_0  &\simeq \frac{135}{8\pi^7\,\gss}\, \sqrt{\frac{10}{\gs}}\,\frac{\Mpl\, \Trh^{3}}{\Lambda^{4}}~,
    \eeq
for instantaneous reheating \cite{Elahi:2014fsa}, where $T_{\rm RH}$ is the highest temperature of the thermal bath and we assume $m_{\chi}\ll T_{\rm RH}$. For the dimension-six interaction in Eq.~\eqref{eq:freezein-operator}, $n=2$. We note in passing that for non-instantaneous reheating the yield also depends on the details of reheating and on the expansion history \cite{Bernal:2019mhf,Giudice:2000ex}.
Furthermore, an interesting variant is the case of Boltzmann suppressed UV freeze-in scenario with  $m_{\chi}\gtrsim T_{\rm RH}$, in this case (and similarly for IR freeze-in \cite{Cosme:2023xpa})  the final abundance is exponentially suppressed \cite{Bernal:2025fcl}
\beq \label{eq:Yafter-BS}
    Y_0^{\rm BS}  \simeq \frac{135}{8\pi^7\,\gss}\, \sqrt{\frac{10}{\gs}}\,\frac{\Mpl\, \Trh^{3}}{\Lambda^{4}}   \frac{3}{2} \left(\frac{m_{\chi}}{\Trh}\right)^{3} e^{- \frac{2 m_{\chi}}{\Trh}} ~.
\eeq
Notably, Boltzmann suppressed variants can enhance detection prospects, and this is true for the case of mixed PBH-FIMP dark matter (as observed in \cite{Chanda:2025bpl}).

The total and neutrino partial annihilation rates must be distinguished:
\begin{equation}
\langle\sigma v\rangle_{\rm tot}=\langle\sigma v\rangle_{\nu\bar\nu}
+\langle\sigma v\rangle_{\mu^+\mu^-}
+\langle\sigma v\rangle_{\tau^+\tau^-}.
\end{equation}
The total rate determines the annihilation plateau, whereas the neutrino emissivity is proportional to $\langle\sigma v\rangle_{\nu\bar\nu}$.

Matching the present yield to the observed density relates $T_{\rm RH}$, $m_\chi$, and $\Lambda$. Freeze-in additionally requires
$\Gamma_{\rm SM\rightarrow DM}(T)<H(T)$ (for all $T$) including all relevant initial states and inverse processes throughout reheating and radiation domination.

In the heavy-mediator completion (cf.~Section~\ref{sec6}) the cutoff can be identified with Lagrangian parameters
\begin{equation}
\Lambda^{2}=\frac{m_{Z'}^{2}}{g'^{2}q_\chi}.
\end{equation}
Importantly, EFT validity requires
$m_{Z'}\gg\max(m_\chi,T_{\rm RH},T_{\rm max}),$ and $g' <\sqrt{4\pi}$.

A thermal hidden-sector distribution is obtained only if elastic and number-changing interactions are sufficiently rapid. We therefore require
\begin{equation}
\Gamma_{\rm self}(T')>H(T)
\end{equation}
during internal equilibration. If this condition fails, the PBH halo must be computed from the nonthermal freeze-in phase-space distribution. If the hidden sector is initially cold and subsequently heated by the visible sector, a useful estimate is \cite{Cheung:2010gj}
\beq
\frac{T'_{{\rm FI}}}{T_{{\rm FI}}} \sim \Big(M_{\rm Pl}  \langle \sigma v \rangle_{{\rm FI}} T_{{\rm FI}} \Big)^{1/4}~,
\eeq
where the production cross section is evaluated during freeze-in. Under the additional assumption of rapid hidden-sector kinetic equilibration, we characterize the initial velocity dispersion by $T_{\rm RH}'$ and require that its phase-space moments reproduce those used in the halo calculation. Without internal thermalization, the nonthermal momentum distribution must be evolved directly.

\begin{figure}[t]
\centering{
\includegraphics[scale = 0.6]{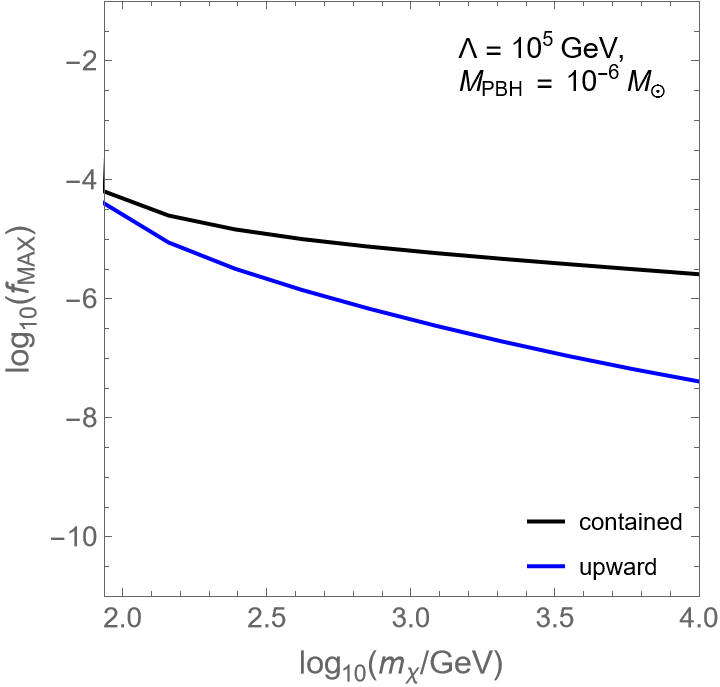} 
%\hspace{1cm}
\includegraphics[scale = 0.595]{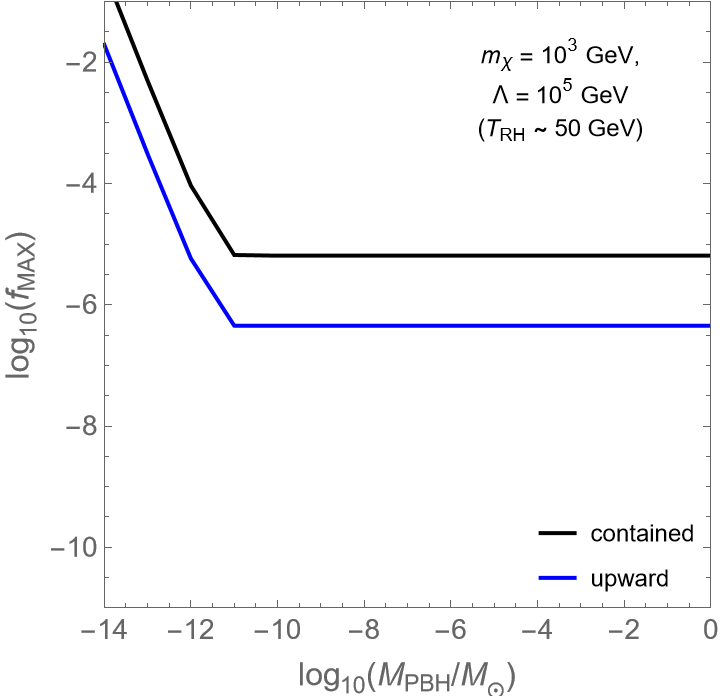}
   \caption{Idealized IceCube freeze-in sensitivities for the mixed FIMP-PBH scenario, assuming dimension-six UV freeze-in in Eq.~\eqref{eq:freezein-operator}, with instantaneous reheating and in the Boltzmann-suppressed regime. Left: $f_{\rm PBH}$ versus $m_\chi$ for $M_\bullet=10^{-6}M_\odot$ and $\Lambda=10^5~{\rm GeV}$. Right: $f_{\rm PBH}$ versus $M_\bullet$ for $m_\chi=10^3~{\rm GeV}$, $\Lambda=10^5~{\rm GeV}$, and $T_{\rm RH}\simeq50~{\rm GeV}$}
      \label{fig:fi-event}}
      \vspace{3mm}
\end{figure}

\begin{figure}[t!]
\centering{
\includegraphics[scale = 0.6]{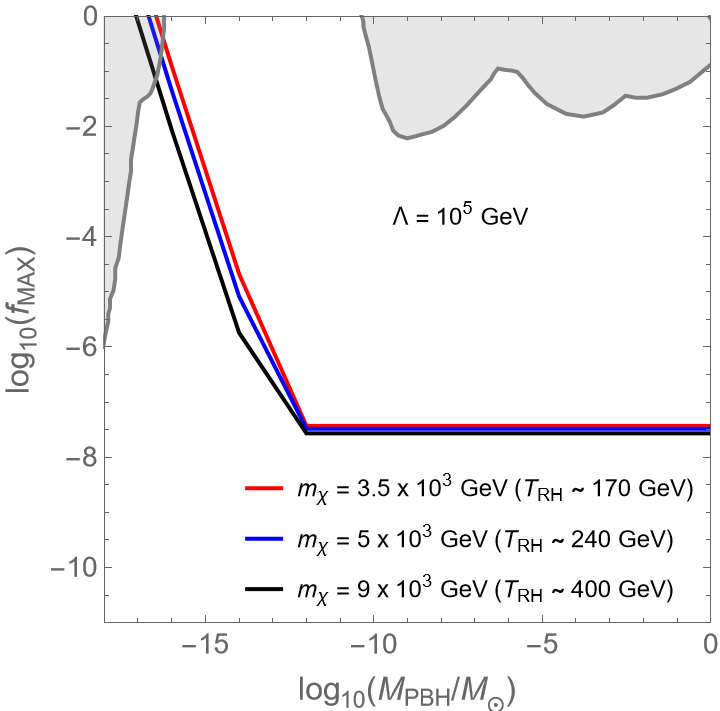} 
%\hspace{1cm}
\includegraphics[scale = 0.6]{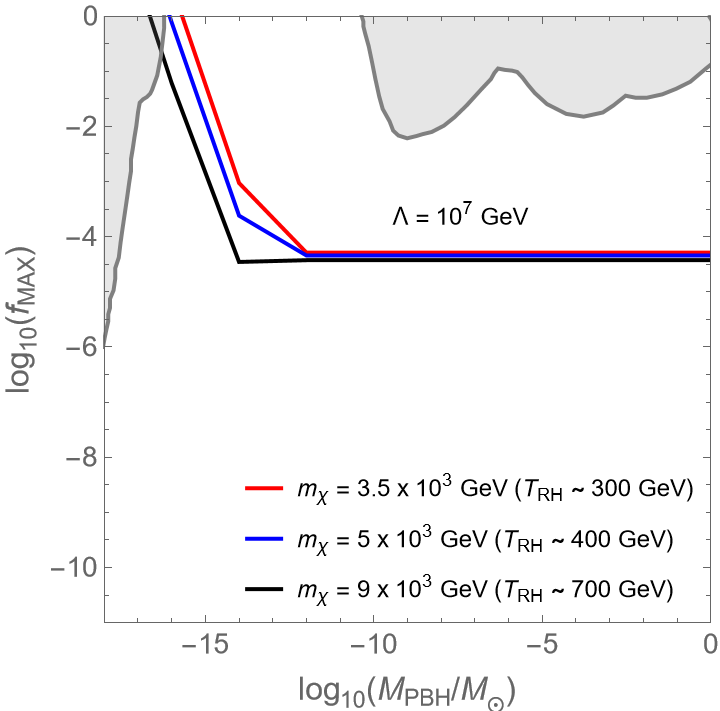}
   \caption{Conservative freeze-in diffuse-flux envelope bounds for the dimension-six operator. The curves use $m_\chi=(3.5,5,9)\times10^3~{\rm GeV}$ and relic-density-matched $T_{\rm RH}$ values shown in the panels. The left and right panels take $\Lambda=10^5~{\rm GeV}$ and $10^7~{\rm GeV}$, respectively. The grey shaded region shows the parameter regions excluded by standard searches for PBH (see, e.g.~\cite{Carr:2020xqk}).}
   \label{fig:fi-flux}}
\end{figure}

Figure~\ref{fig:fi-event} shows the supplied event-based projections, while Figure~\ref{fig:fi-flux} shows the conservative diffuse-flux comparison. The selected benchmarks lie in the Boltzmann-suppressed regime, $m_\chi>T_{\rm RH}$, where relic-density matching requires a larger coupling than in unsuppressed freeze-in \cite{Chanda:2025bpl}. A meaningful comparison with gamma-ray constraints requires the same operator, reheating history, relic-density prescription, branching fractions, and halo model. We therefore postpone a quantitative neutrino-versus-gamma comparison until these inputs are implemented consistently in both channels.

\section{Concluding Remarks}
\label{sec8}

We have studied the neutrino signal  arising from particle dark matter annihilation in PBH-induced minispikes, emphasizing the impact of a more careful treatment of both the astrophysical environment and the particle physics interpretation. In particular, the use of refined, piecewise halo profiles can significantly impact the annihilation luminosity relative to the commonly adopted single power-law form $\rho \propto r^{-9/4}$. Our main results are idealized IceCube sensitivity projections on the mixed WIMP-PBH (Figure \ref{fig:fo-event}) and FIMP-PBH  (Figure \ref{fig:fi-event})  scenarios, as well as complementary conservative limits derived from comparisons with the observed diffuse neutrino flux  (Figures \ref{fig:fo-flux} \& \ref{fig:fi-flux}, respectively).

More broadly, this work highlights the close interplay between cosmology, gravitation, and particle physics in determining the indirect signatures of dark matter. Primordial black holes provide a natural mechanism for enhancing otherwise suppressed dark matter annihilation signals, making neutrino observations a particularly valuable complement to traditional gamma-ray searches. With forthcoming high-energy neutrino data, these systems may provide one of the most sensitive laboratories for exploring the possible coexistence of primordial black holes and particle dark matter.

\vspace{5mm}
\noindent {\bf Acknowledgments.} JU is supported by NSF grant PHY-2209998. During the final stages of this work, the preprint \cite{Yang:2026zcq} appeared, which also examines neutrino signals from dark matter annihilation around PBH and which partially overlaps our study. The two works can be viewed as complementary analyses.

%%%%%%%%%%%%%%%%%%%%%%%%%% Appendix %%%%%%%%%%%%%%%%%%%%%%%%%%%%%%%%%%%%%% 
 \appendix

 \section{Dark-matter halo profiles around PBHs}
\label{sec:spike-density}

While the density profile $\rho\propto r^{-9/4}$ provides a characteristic estimate, the halos of dark matter particle around PBHs tend to be significantly more complicated. Accordingly, the halo profiles must be computed carefully. A careful study of halo profiles around PBH has been undertaken in Boudaud {\em et al.} \cite{Boudaud:2021irr}, although neglecting particle annihilation within the halo. Dark matter annihilation effects were incorporated later in \cite{Chanda:2022hls}, where it was emphasized that, for WIMP-like candidates, annihilations can substantially reshape the late-time inner profile. (See also recent analytic developments in \cite{Lavalle:2025rnx}.)
This Appendix compiles the key halo-profile formulas reported by Boudaud {\em et al.} \cite{Boudaud:2021irr} and \cite{Chanda:2022hls}, which we use in Section \ref{sec2}.

For the basic formation picture, we begin with the radial equation of motion for a thin shell of particles at distance $r$ from the PBH (see, for example, \cite{Adamek:2019gns}):
\begin{equation}
    \ddot{r} = -\frac{G M_{\bullet}}{r^2} + \frac{\ddot{a}}{a}r~,
\end{equation}
The first term is simply the Newtonian pull from the PBH, while the second term captures the background deceleration of an FLRW Universe during radiation domination. When the magnitudes of these two contributions are comparable, one reaches the point of turnaround.
The turnaround radius is the distance from the PBH at a given time $t$ for which a dark matter shell stops following the Hubble flow due to the PBH's gravitational pull (see e.g.~\cite{Adamek:2019gns})
\beq \label{rta}
r_{{\rm ta}}(t)\simeq (2 G M_{\bullet}  t^2)^{1/3}~.
\eeq
The cosmic time can be related to the temperature  (during radiation domination) via the standard relation
 \beq
 \label{tT}
 t = \frac{1}{T^2}\sqrt{\frac{45}{16 \pi^3 G g_*}}~.
 \eeq  
 By construction, $r_{\rm ta}(t)$ is the radius of the shell that has just detached from the Hubble expansion and begun infall at time $t$.

In the framework of \cite{Boudaud:2021irr}, the halo density can be expressed in terms of the dimensionless radius $\tilde{r}=r/r_s$, where $r_s=2GM_{\bullet}$ is the Schwarzschild radius, as  \cite{Boudaud:2021irr}
\begin{equation} \label{eq:rhor}
    \rho(\tilde{r}) = \sqrt{\frac{2}{\pi^3}} \frac{\rho_i}{\sigma^3_i} \tilde{r}^{-3/2} \int \int {\rm d} \mathcal{R} {\rm d}u \left\{ \mathcal{R}(1-u) \right\}^{3/2} \Theta (\bar{u}(\tilde{r})-u) \int_{\sqrt{\mathcal{Y}_m} \Theta(\mathcal{Y}_m)}^1 \frac{{\rm d}y}{\sqrt{y^2 - \mathcal{Y}_m}}~,
\end{equation}
with $\bar{u}(\tilde{r})=\tilde{r} \sigma_{\rm collapse}^2(\tilde{r}) $ and $\mathcal{Y}_m = 1 + \mathcal{R}^2 \left\{ \frac{1}{u} \left(1 - \frac{1}{\mathcal{R}}\right) - 1 \right\}$. 
Computing the integral in eq.~(\ref{eq:rhor}) yields three analytic branches, corresponding to the following inequalities:
\begin{itemize}[leftmargin=0.75in]
 \item[(i).] ~~$\bar{u}(\tilde{r}_{\rm eq}) < \bar{u}(\tilde{r}_{i}) < 1$, 
 \item[(ii).] ~~$\bar{u}(\tilde{r}_{\rm eq}) < 1 < \bar{u}(\tilde{r}_{i})$, 
 \item[(iii).] ~~$1 < \bar{u}(\tilde{r}_{\rm eq}) < \bar{u}(\tilde{r}_{i})$.
 \end{itemize}
It is useful to rewrite these cases as conditions on $M_{\bullet}$ relative to two reference masses,
\begin{equation}
    M_1 = \left( \frac{T_i}{m_{\rm DM}} \right)^{3/2} \frac{\eta_{\rm ta}^{1/2} t_i }{2G}~,
\end{equation}
and 
\begin{equation}
    M_2 =  \left( \frac{T_i}{m_{\rm DM}} \right)^{3/2} \frac{\eta_{\rm ta}^{1/2} t_{\rm eq} }{2G} \left( \frac{T_{\rm eq}}{T_i} \right)^3~.
\end{equation}
where $T_i$ is the temperature of the dark matter at the formation time $t_i$.
Following \cite{Boudaud:2021irr}, one then has three categories, corresponding to heavy, intermediate, and light PBHs. In each regime the density is described by a hierarchy of power-law pieces, summarized below. These expressions describe the profile prior to including dark matter microphysics (e.g. annihilation) and before accounting for tidal stripping from close encounters. Additional details and derivations can be found in \cite{Boudaud:2021irr,Chanda:2022hls}.

\subsection{Heavy PBH regime}
In the heavy-PBH ordering $M_{\bullet} > M_1 > M_2$, the profile takes the form
\begin{equation}
    \rho(\tilde{r}) = \begin{cases}
        \rho_{3/2}(\tilde{r}) &~~~~~ 0<\tilde{r}<\tilde{r}_1, \\
        \rho_{9/4}(\tilde{r}) &~~~~~ \tilde{r}_1<\tilde{r}<\tilde{r}_{\rm eq}, \\
        0 &~~~~~ \tilde{r}_{\rm eq}<\tilde{r},
    \end{cases}
\end{equation}
where
\begin{equation}
    \rho_{3/2}(\tilde{r})=\sqrt{\frac{2}{\pi^3}} \rho_{i}\tilde{r}^{3/2}_{i}\left[ \frac{2 \sqrt{2 \pi}}{3} \left\{2 + \left(1+2\frac{\tilde{r}}{\tilde{r}_{i}}\right) \sqrt{1-\frac{\tilde{r}}{\tilde{r}_{i}}}\right\}\right]\tilde{r}^{-3/2}~,
\end{equation}
and
\begin{equation}
    \rho_{9/4}(\tilde{r})= \frac{\sqrt{128 \pi}}{\Gamma^2(\frac{1}{4})} \rho_{\rm eq}\tilde{r}_{\rm eq}^{9/4} \left\{1-\frac{\Gamma^2(\frac{1}{4})}{3\sqrt{2\pi^3}}\left(\frac{\tilde{r}}{\tilde{r}_{\rm eq}}\right)^{3/4}\right\}\tilde{r}^{-9/4}~,
\end{equation}
with $\rho_{\rm eq} \equiv \rho_{\rm collapse}(\tilde{r}_{\rm eq})$. The matching radius $\tilde{r}_1$ is fixed by enforcing continuity of $\rho(\tilde{r})$ at $\tilde{r}=\tilde{r}_1$.

\subsection{Intermediate PBH regime}
When $M_1 > M_{\bullet} > M_2$, the halo again consists of nested power laws,
\begin{equation}
    \rho(\tilde{r}) = \begin{cases}
        \rho_{3/4}(\tilde{r}) &~~~~~ 0<\tilde{r}<\tilde{r}_1 \\
        \rho'_{3/2}(\tilde{r}) &~~~~~ \tilde{r}_1<\tilde{r}<\tilde{r}_2 \\
        \rho_{9/4}(\tilde{r}) &~~~~~ \tilde{r}_2<\tilde{r}<\tilde{r}_{\rm eq} \\
        0 &~~~~~ \tilde{r}_{\rm eq}<\tilde{r}
    \end{cases}~.
\end{equation}
In the innermost region one has $\rho\propto\tilde{r}^{-3/4}$, with
\begin{equation}
    \rho_{3/4}(\tilde{r}) \simeq 4.2 \sqrt{\frac{2^{5/2}}{\pi^3}}  \Gamma\left(\frac{7}{4}\right) \frac{\rho_{i}}{\sigma_{i}^{3/2}}\tilde{r}^{-3/4}~.
\end{equation}
At large radii the scaling is $\rho\propto \tilde{r}^{-9/4}$, matching the heavy-PBH case. Between these regions there is an intermediate scaling $\rho\propto \tilde{r}^{-3/2}$; here the functional form differs from the heavy case (hence the prime) and reads
\begin{equation}
    \rho'_{3/2}(\tilde{r}) = \sqrt{\frac{2}{\pi^3}} \frac{\rho_{i}}{\sigma^3_{i}} \left(1.047-\frac{3 \pi}{8} \frac{\tilde{r}}{\tilde{r}_{\rm eq}} \right) \tilde{r}^{-3/2}~,
\end{equation}
The radii $\tilde{r}_1$ and $\tilde{r}_2$ are determined by matching the adjoining pieces at each boundary and solving for the intersections.

\subsection{Light PBH regime }
In the light-PBH ordering $M_1 > M_2 > M_{\bullet}$, the profile simplifies to
\begin{equation}
    \rho(\tilde{r}) = \begin{cases}
        \rho_{3/4}(\tilde{r}) &~~~~~ 0<\tilde{r}<\tilde{r}_1 \\
        \rho'_{3/2}(\tilde{r}) &~~~~~ \tilde{r}_1<\tilde{r}<\tilde{r}_{\rm eq} \\
        0 &~~~~~ \tilde{r}_{\rm eq}<\tilde{r}
    \end{cases}~.
\end{equation}
The functions $\rho_{3/4}$ and $\rho'_{3/2}$ are exactly those given in the intermediate regime. The single transition scale $\tilde{r}_1$ is again obtained by continuity at the matching point.

\section{Origin of the three regimes in the $f_{\rm PBH}$-$m_\chi$ limits}
\label{ApD}

The projected upper bound on the PBH fraction for the contained event is determined by the ratio of the signal and background event rates
\begin{align}
f_{\rm PBH}^{\rm lim}
\propto
\frac{\sqrt{N_{\rm atm}}}
     {N_{\rm signal}}.
\end{align}
The characteristic shape of the $f_{\rm PBH}$-$m_\chi$ curve arises from three competing effects.
At low dark matter masses, the neutrino energy $E_\nu \simeq m_\chi$ lies close to the detector threshold. The neutrino-nucleon interaction cross section, muon range, and detector acceptance are all suppressed, leading to a small signal rate and consequently weak limits on $f_{\rm PBH}$.

As $m_\chi$ increases, the detector efficiency improves while the atmospheric neutrino background decreases rapidly with energy, approximately as
$\Phi_{\rm atm}\propto E_\nu^{-\gamma}$, where $\gamma$ parameterises a power-law.
This enhances the signal-to-background ratio and produces the strongest sensitivity at intermediate masses.
For sufficiently large $m_\chi$, the annihilation rate around each PBH decreases because the dark matter number density scales as $n_\chi=\rho_\chi/m_\chi$. Since
\begin{align}
\Gamma_{\rm ann}
\propto
\int \frac{\rho_\chi^2}{m_\chi^2}
\langle \sigma v\rangle\, dV,
\end{align}
the signal rate eventually falls with increasing $m_\chi$. This suppression overcomes the continuing reduction of the atmospheric background, causing the $f_{\rm PBH}$ limits to weaken again at high~$m_\chi$.

Therefore, the turnover in the $f_{\rm PBH}$-$m_\chi$ plane is a direct consequence of:
\begin{enumerate}
\item detector-threshold suppression at low energies,
\item rapidly falling atmospheric backgrounds at intermediate energies,
\item annihilation-rate suppression due to decreasing dark matter number density at large~$m_\chi$.
\end{enumerate}

The transitions between the light-, intermediate-, and heavy-PBH halo regimes occur at $M_\bullet=M_2$ and $M_\bullet=M_1$.  The approximately flat portions of the $M_\bullet$ constraints arise from the compensating scalings of the annihilation luminosity per minispike and the PBH number density within a fixed profile branch.

\end{document}